\documentclass[
    ,final            
  ]
  {aipproc}

\layoutstyle{6x9}

\begin{document}

\title{Status of Longitudinal Polarized Parton \\[2mm]
Densities and Higher Twist}



\author{E. Leader}{
  address={Imperial College London, London WC1E 7HX, England}
}

\author{A.V. Sidorov,}{
  address={Bogoliubov Theoretical Laboratory,
Joint Institute for Nuclear Research, 141980 Dubna, Russia } }

\author{\underline{D.B. Stamenov}
\thanks{This research was supported by the JINR-Bulgaria Collaborative
Grant, by the RFBR (No 07-02-01046, 08-01-00686) and by the
Bulgarian National Science Foundation under Contract 288/2008.}
} {
  address={Institute for Nuclear Research and Nuclear Energy,
Bulgarian Academy of Science, Sofia 1784, Bulgaria}
}

\begin{abstract}
The present status of the longitudinal polarized parton densities
(PDFs) and the contribution of their first moments to the nucleon
spin is discussed. Special attention is paid to the role of higher
twist effects in determining the PDFs and to the polarized strange
quark and gluon densities, which are still not well determined
from the present data.
\end{abstract}

\maketitle


\section{Introduction and Methods of Analysis}

Our present knowledge about the nucleon spin structure comes
mainly from polarized inclusive and semi-inclusive DIS experiments
at SLAC, CERN, DESY and JLab, polarized proton-proton collisions
at RHIC and polarized photoproduction experiments. One of the
important and best studied aspects of this knowledge is the
determination of the longitudinal polarized parton densities in
QCD and their first moments, which correspond to the spins carried
by the quarks and gluons in the nucleon.

There are some peculiarities which make the QCD analysis of the
data more complicated and difficult than that in the unpolarized
case:

~i) A half of the present polarized DIS data are at moderate $Q^2$
and $W^2~(Q^2 \sim 1-4~\rm GeV^2,~4~\rm GeV^2 < W^2 < 10~\rm
GeV^2$), or in the so-called preasymptotic region. So, in contrast
to the unpolarized case, the $1/Q^2$ terms ({\it kinematic} -
$\gamma^2$ factor, target mass corrections, and {\it dynamic} -
higher twist corrections to the spin structure function $g_1$)
have to be accounted for in the analysis in order to determine
{\it correctly} the polarized PDFs.

ii) Due to the lack of charged current neutrino data, only the sum
$(\Delta q + \Delta\bar{q})$ can be determined from inclusive DIS
and information from semi-inclusive DIS (SIDIS) is needed for the
flavor decomposition of the polarized sea ($\Delta\bar{u},
~\Delta\bar{d},~\Delta\bar{s}$). Such an analysis, however,
requires knowledge of the fragmentation functions which are not
well known at present. Note that in the unpolarized case, SIDIS
processes have never been used for a flavor separation of the
parton densities.

The best manner to determine the polarized PDFs is to perform a
QCD fit to the data on $g_1/F_1$, which can be obtained if both
$A_{||}$ and $A_{\perp}$ asymmetries are measured. In the case if
only $A_{||}$ is measured, the quantity $A_{||}/D(1+\gamma^2)$ is
a good approximation of $g_1/F_1$, where D is the depolarization
factor and $\gamma^2=4M^2x^2/Q^2$. The data on the photon-nucleon
asymmetry $A_1$ are not suitable for the determination of PDFs
because the structure function $g_2$ is not well known in QCD and
the approximation $(A_1)^{theor}(= g_1/F_1-\gamma^2g_2/F_1)\approx
(g_1/F_1)^{theor}$ in the preasymptotic region used by some of the
groups is {\it not} reasonable.

In QCD, one can split $g_1$ and $F_1$ into leading (LT) and higher
twist (HT) pieces
\begin{equation}
g_1 = (g_1)_{\rm LT} + (g_1)_{\rm HT},~~~~F_1 = (F_1)_{\rm LT} +
(F_1)_{\rm HT}. \label{LTHT}
\end{equation}
Then, approximately
\begin{equation}
{g_1\over F_1} \approx {(g_1)_{\rm LT}\over (F_1)_{\rm LT}}[1 +
{(g_1)_{\rm HT}\over (g_1)_{\rm LT}} - {(F_1)_{\rm HT}\over
(F_1)_{\rm LT}}]. \label{g1ovF1}
\end{equation}
Note that LT pieces of $g_1$ and $F_1$ are expressed by the
polarized and unpolarized PDFs, respectively.

There are mainly two methods to fit the data - taking or NOT
taking into account the HT corrections to $g_1$. According to the
first \cite{LSS_HT}, the data on $g_1/F_1$ have been fitted taking
into account the contribution of the first term $h(x)/Q^2$ in
$(g_1)_{\rm HT}$ and using the experimental data for the
unpolarized structure function $F_1$
\begin{equation}
\left[{g_1(x, Q^2)\over F_1(x, Q^2)}\right]_{exp}~\Leftrightarrow~
{{g_1(x,Q^2)_{\rm LT}+h(x)/Q^2}\over F_1(x,Q^2)_{exp}}~.
\label{g1HTf1}
\end{equation}
According to the second approach \cite{GRSV} only the LT
expression in (\ref{LTHT}) have been used in the fit to the
$g_1/F_1$ data
\begin{equation}
\left[{g_1(x, Q^2)\over F_1(x, Q^2)}\right]_{exp}~\Leftrightarrow~
{{g_1(x,Q^2)_{\rm LT}}\over F_1(x,Q^2)_{\rm LT}}~. \label{g1f1_LT}
\end{equation}

It is obvious that the two methods are equivalent in a pure DIS
region where HT can been ignored. To be equivalent in the {\it
preasymptotic} region requires a cancellation between the ratios
$(g_1)_{\rm HT}/(g_1)_{\rm LT}$ and $(F_1)_{\rm HT}/(F_1)_{\rm
LT}$ in (\ref{g1ovF1}). Fig. 1 (left), based on our recent results
on $g_1$ \cite{LSS06} and the unpolarized results of
\cite{Alekhin}, demonstrates the validity of this for $x\geq
0.15$, but clearly indicates that ignoring HT terms in the ratio
${g_1}/{F_1}$ below $x=0.15$, as was done by groups using the
second method or some its modifications \cite{groups, COMPASS}, is
incorrect. Note that in this case the HT effects are absorbed in
the extracted PDFs which differ from those determined in the
presence of HT (see \cite{LSS06} for more details).
\begin{figure}
  \includegraphics[height=.29\textheight]{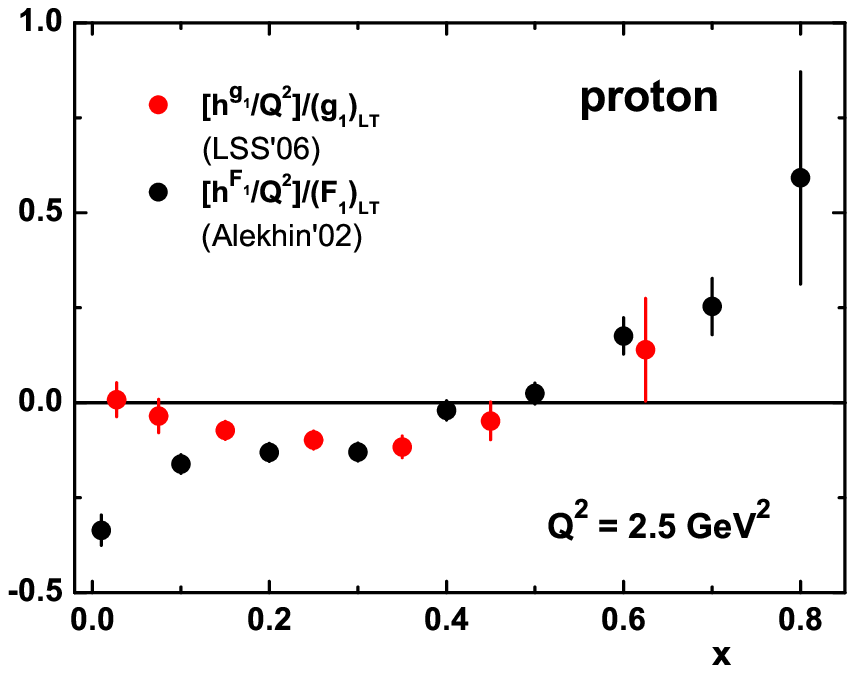}
    \includegraphics[height=.29\textheight]{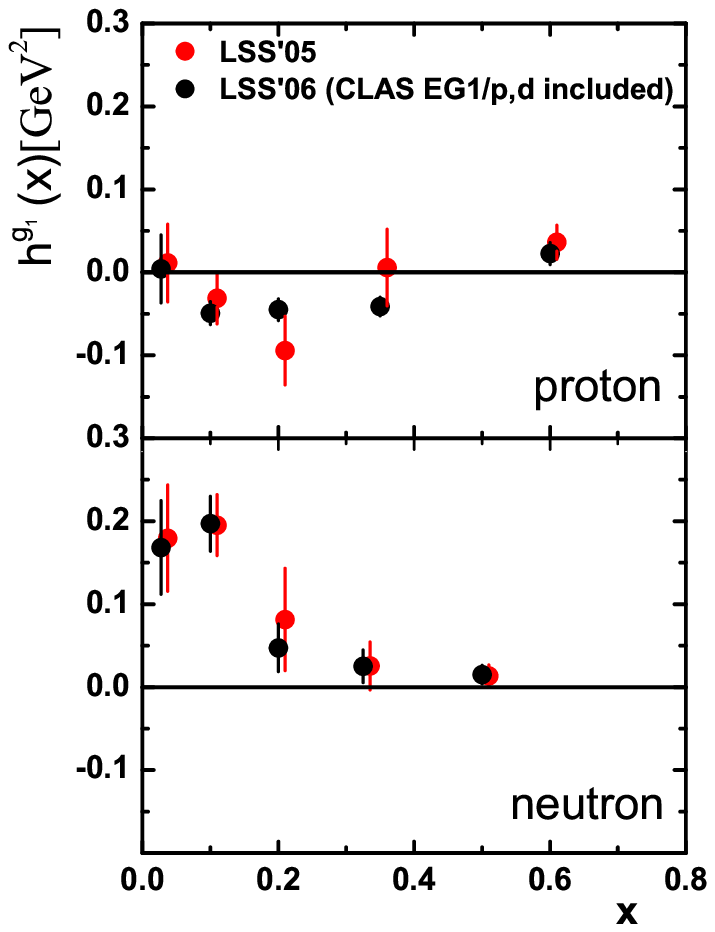}
  \caption{Comparison of HT terms in $g_1$ and $F_1$ (left).
HT effects for protons and neutrons (right).}
\end{figure}

\section{Polarized Parton Densities and Higher Twist}

Let us start our discussion with the impact of the COMPASS
\cite{COMPASS} and CLAS \cite{CLAS} inclusive DIS data on the {\it
longitudinal} polarized PDFs and higher twist effects. The precise
CLAS data on $g_1^p/F_1^p$ and $g_1^d/F_1^d$ are data at low
$Q^2(1-4~GeV^2)$ while COMPASS has presented data on $A_1^d$
mainly at large $Q^2$ and very small $x~(0.004\leq x \leq 0.02)$,
the \emph{only} precise data at such small $x$. Compared to the HT
values obtained in the LSS'05 analysis \cite{LSS05}, the
uncertainties in the HT values at each $x$ in the CLAS region ($x
\sim 0.1-0.6$) are significantly reduced by the CLAS data, as seen
in Fig. 1 (right). As expected, the central values of the
polarized PDFs are practically {\it not} affected by the CLAS
data. This is a consequence of the fact that at low $Q^2$ the
deviation from logarithmic in $Q^2$ pQCD behaviour of $g_1$ is
accounted for by the higher twist term in $g_1$. The accuracy of
the determination of polarized PDFs, however, is essentially
improved.
\begin{figure}
  \includegraphics[height=.35\textheight]{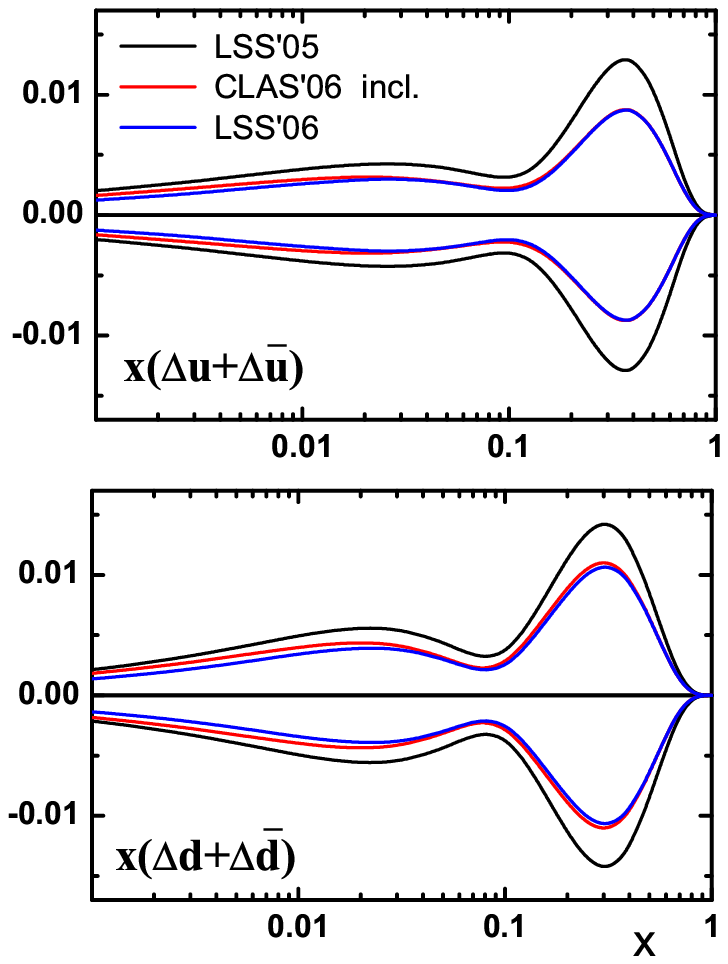}
    \includegraphics[height=.35\textheight]{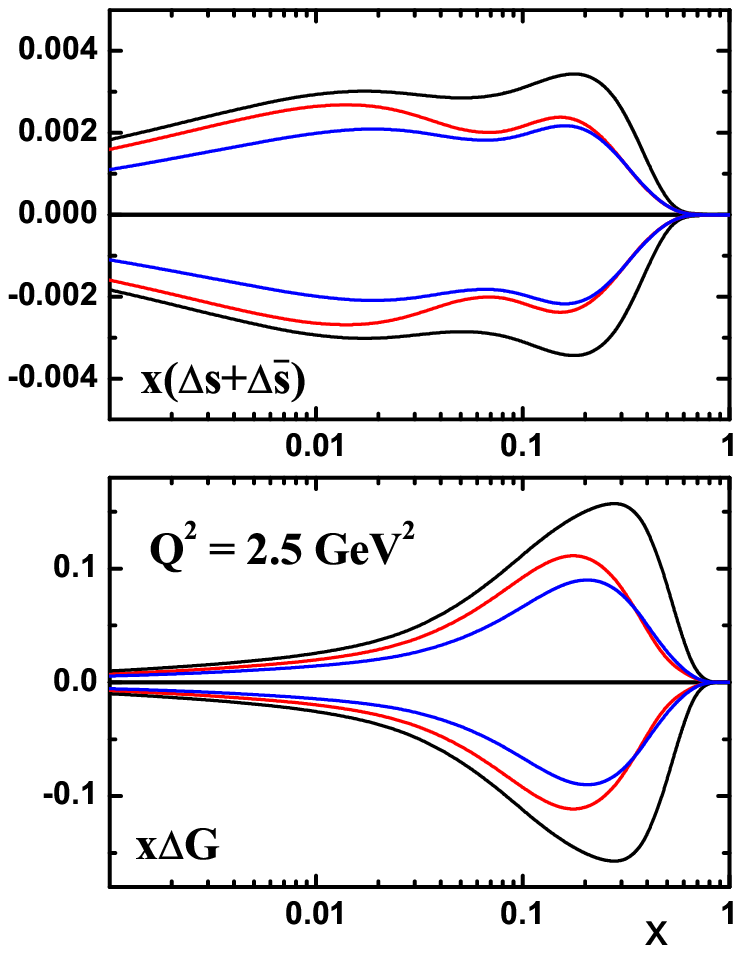}
  \caption{Impact of CLAS and COMPASS data on the uncertainties for
  NLO($\rm \overline{MS}$)
  polarized PDFs.}
\end{figure}
This improvement is illustrated in Fig. 2 (red curves). It is a
consequence of the much better determination of higher twist
contribution in $g_1$, as discussed above. The impact of COMPASS
data on the PDFs uncertainties is also shown in Fig. 2 (the blue
curves). As seen, they help to improve in addition the accuracy of
the determination of the gluon and strange sea quark polarized
densities at $x<0.2$ and $x<0.1$, respectively. Another important
effect of COMPASS data (see \cite{LSS06} and slides of Stamenov's
talk at this Symposium \cite {slides}) is that $\Delta s(x)$ and
respectively, $\Delta \Sigma(x)$, are less negative at x smaller
than 0.1.

The present {\it inclusive} DIS data cannot distinguish between
positive, negative and sign-changing $\Delta G(x)$ (see Fig. 3
(left)). In all the cases the magnitude of $\Delta G$ is small:
$|\Delta G| < 0.4$ and the corresponding polarized quark densities
are very close to each other. COMPASS analysis finds also a
negative $\Delta G(x)$, but has some peculiarities which suggest
it is not physical. Also, we have found that if in the input gluon
parametrization a term $(1+ \gamma x)$ is added, where $\gamma$ is
a free parameter to be determined from the fit to the data, a
negative $\Delta G$ solution is impossible. So, although the
negative $\Delta G$ cannot be formally ruled out by the data, it
seems to be {\it not} reasonable.
\begin{figure}
  \includegraphics[height=.27\textheight]{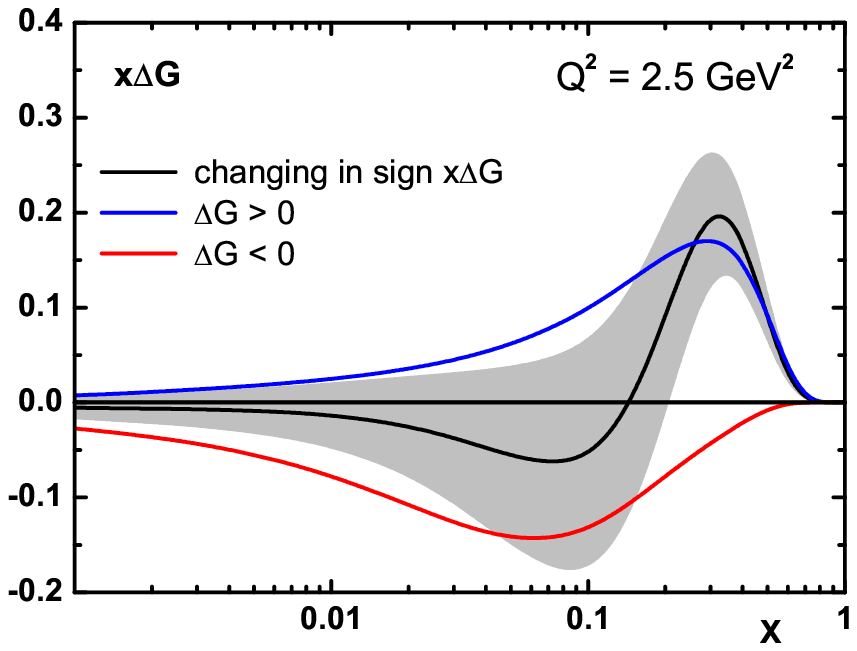}
    \includegraphics[height=.27\textheight]{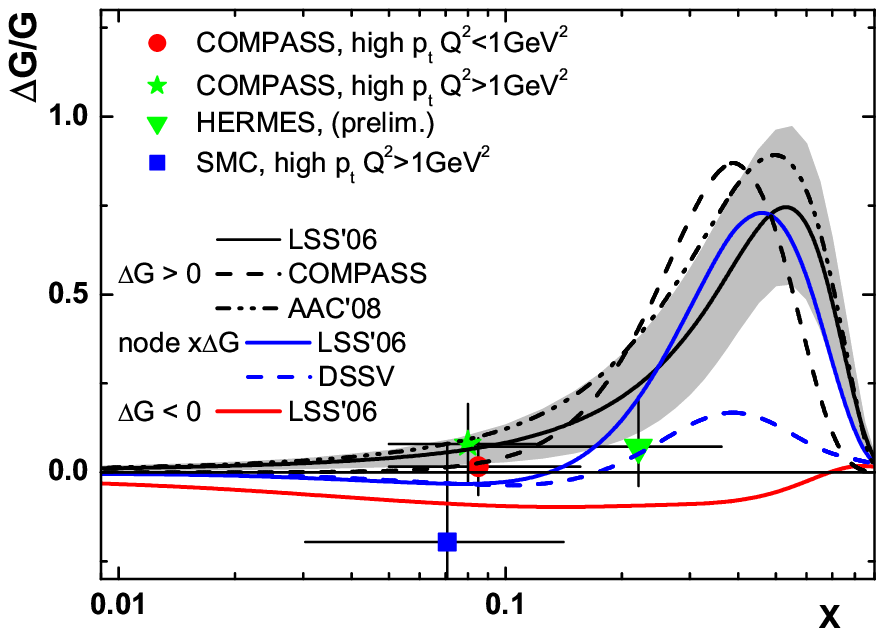}
  \caption{The three different LSS'06 solutions for $\Delta G(x)$ (left).
  Comparison with directly measured $\Delta G(x)/G(x)$ (right).}
\end{figure}

A primary goal of the RHIC spin program is to determine the gluon
polarization. There are two sources of information:
$\vec{p}+\vec{p} \rightarrow \pi^0(\pi^{+/-}) +
X,~~\vec{p}+\vec{p} \rightarrow jet + X.$ We will mention here the
combined QCD analysis of the inclusive DIS and RHIC
$\pi^0$-production data \cite{PHENIX} which has been recently
performed by AAC \cite{AAC08}. Two solutions for $\Delta G(x)$,
positive and node, have been found  with values for the first
moments at $Q^2= 1~GeV^2:\Delta G$ = 0.40 +/- 0.28   and   -0.12
+/- 1.78, respectively. Note that for the node $\Delta G(x)$
solution the positivity condition $\Delta G(x) \leq G(x)$ is
broken for $x \geq 0.4$, which makes it questionable. For more
details about the possibilities of the RHIC data to constrain the
gluon polarization $\Delta G$ see the talks presented by STAR and
PHENIX Collaborations at this Symposium \cite{slides}. The main
conclusion is that the magnitude of $\Delta G$ is smaller than
0.4, but the present RHIC data cannot determine the form of the
gluon density $\Delta G(x)$.

In Fig. 3(right) we compare the polarized gluon densities obtained
by different QCD analyses (divided by the MRST'02 version of the
unpolarized gluon density $G(x)$) with directly measured values of
$\Delta G(x)/G(x)$ at $\mu^2 = 3~GeV^2$. One can see from Fig.
3(right) that the most precise values of $\Delta G/G$, the COMPASS
ones \cite{dGovG}, are well consistent with most of the polarized
gluon densities determined in  the recent QCD analysis, and one
conclude that the present high $p_t$ measurements cannot
distinguish between the different scenarios for $\Delta G(x)$.

It seems clear that present day data cannot distinguish between
the three scenarios for $\Delta G(x)$. A clean distinction would
be possible in the Electron-Ion Collider \cite{EIC} where very
precise measurements of $g_1^p(x, Q^2)$ at very small $x~(x \geq
0.00075)$ could be performed. Fig. 4 shows $g_1^p(x, Q^2)$,
calculated using the LSS'06 PDFs corresponding to positive and
negative $\Delta G(x)$. There is a dramatic difference at small
$x$.
\begin{figure}
  \includegraphics[height=.24\textheight]{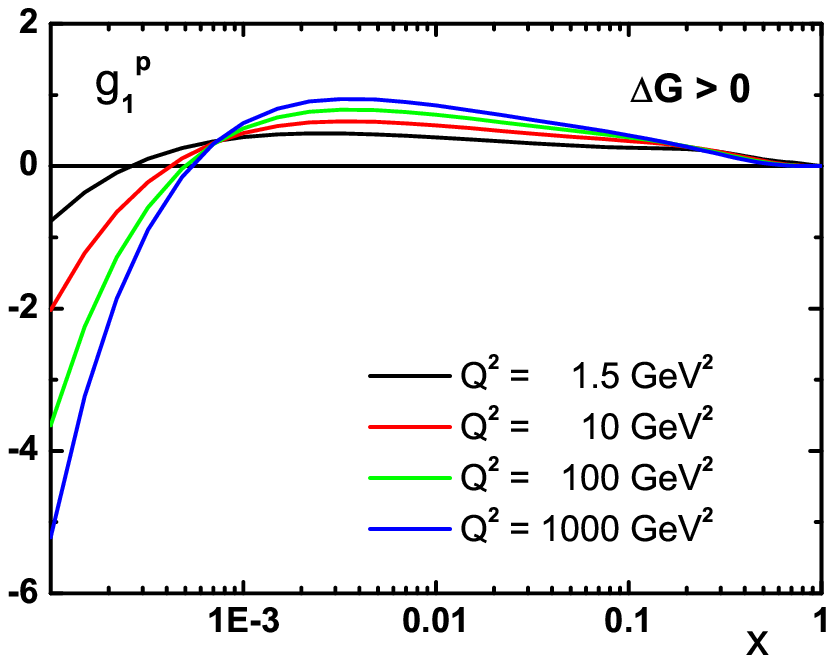}
    \includegraphics[height=.24\textheight]{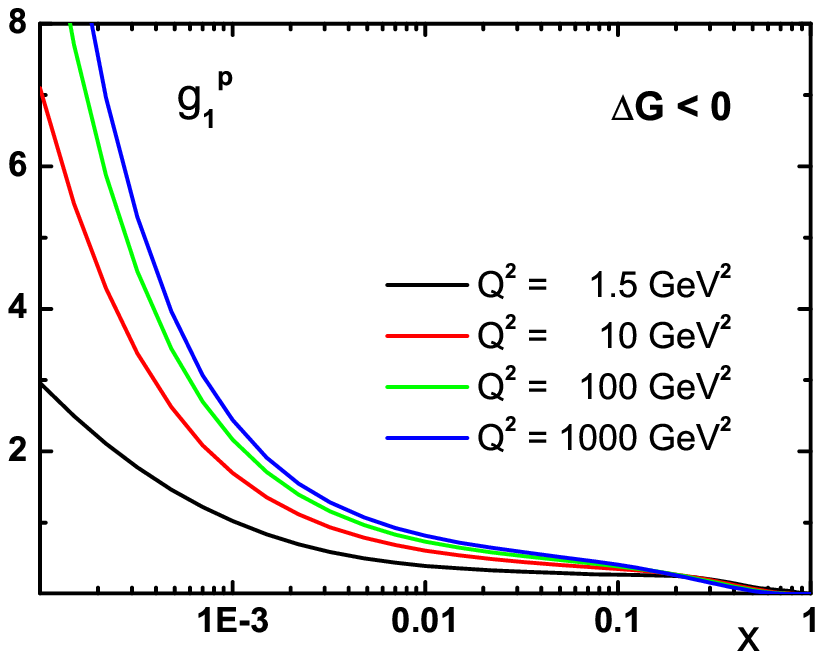}
  \caption{$g_1^p$ at different $Q^2$ calculated using the
LSS'06 PDFs corresponding to $\Delta G > 0$ and $\Delta G < 0$.}
\end{figure}

Very recently the DSSV group has presented results on polarized
PDFs \cite{DSSV} obtained from the first global analysis of the
polarized DIS, SIDIS and RHIC polarized pp scattering data. Due to
the SIDIS data a flavor decomposition of the polarized sea is
achieved. This analysis yields changing in sign $\Delta \bar{s}(x,
Q^2)$: positive for x > 0.03 and negative for small x (see Fig.
5). Its first moment is negative (fixed in practice by the SU(3)
symmetric value of $a_8$) and almost identical with that obtained
in the inclusive DIS analysis. It was shown in the talk by R.
Windmolders at this Symposium \cite{slides} that the determination
of $\Delta \bar{s}(x)$ from SIDIS strongly depends on the
fragmentation functions (FFs) and that for such an unexpected
behavior of $\Delta \bar{s}(x)$ the new FFs \cite{FF} are crucial.
So, the model independent extraction of FFs is very important. The
NLO($\rm \overline{MS}$) PDFs determined from analyses of
different sets of data: LSS'06 (inclusive DIS data), AAC'08
(iclusive DIS and RHIC $\pi^0$-production data), DSSV (DIS, SIDIS
and RHIC data) are compared in Fig. 5. Note that in the LSS
analysis the HT corrections have been taken into account while
AAC'08 and DSSV have used the second method described above
(without account for HT terms). Although the first moments are
almost identical, the quark densities themselves are different,
especially $\Delta \bar{s}(x)$. Note that all analyses of the
inclusive DIS data yield $\Delta \bar{s}(x)$ negative. So, the
DSSV result on $\Delta \bar{s}(x)$ is a big challenge for our
understanding of spin properties of the nucleon. We would like to
emphasize also that the very accurate DIS data in the
preasymptotic region require a more precise confrontation of QCD
to the data, as was already mentioned in the Introduction.
Otherwise, the results on PDFs will differ. A comment on an
incorrect confrontation of QCD to the inclusive DIS data in the
DSSV analysis is presented in Stamenov's talk \cite{slides} at
this Meeting.

Let us finally discuss the present status of the proton spin sum
rule. Using the values for $\Delta \Sigma(Q^2)$ and $\Delta
G(Q^2)$ at $Q^2=4~GeV^2$ obtained in LSS'06 analysis \cite{LSS06}
one can find for the spin of the proton (the numbers in brackets
correspond to node $\Delta G$):
\begin{equation}
S_z = {1\over 2} = {1\over 2}\Delta \Sigma(Q^2) + \Delta G(Q^2) +
L_z(Q^2) = 0.55(0.15) \pm 0.25(0.49) + L_z(Q^2).
\label{SSR}
\end{equation}
Although the central values of parton contribution are very
different in the two cases, in view of the big uncertainty in
(\ref{SSR}) coming mainly from the gluons, one cannot make a
definite conclusion about the quark-gluon contribution in the spin
of the nucleon.
\begin{figure}
  \includegraphics[height=.35\textheight]{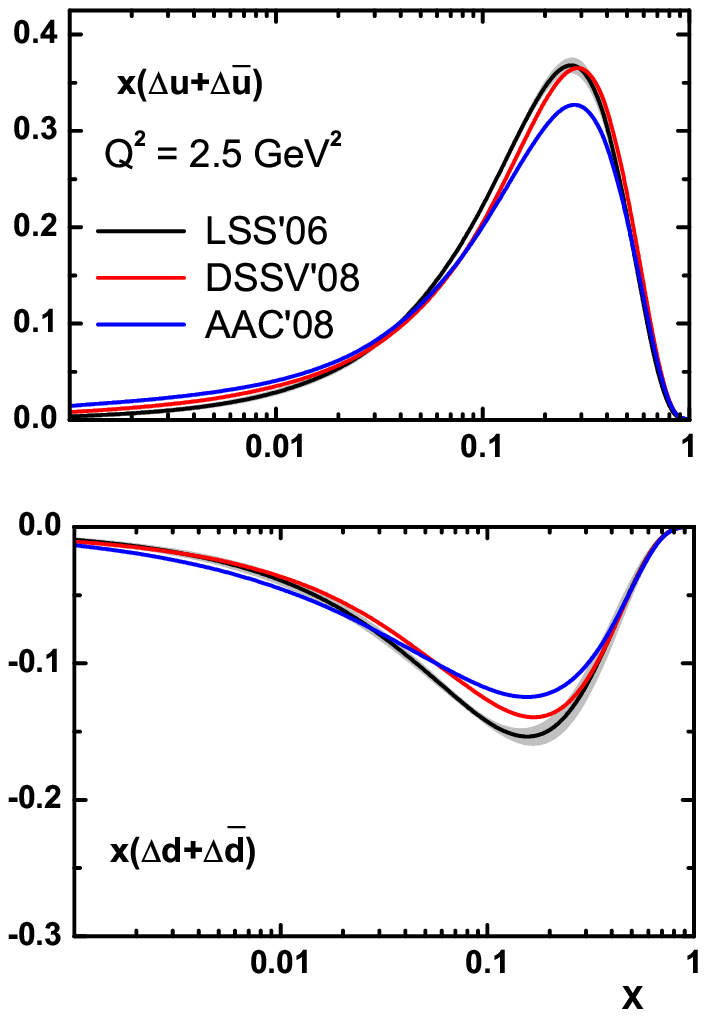}
    \includegraphics[height=.35\textheight]{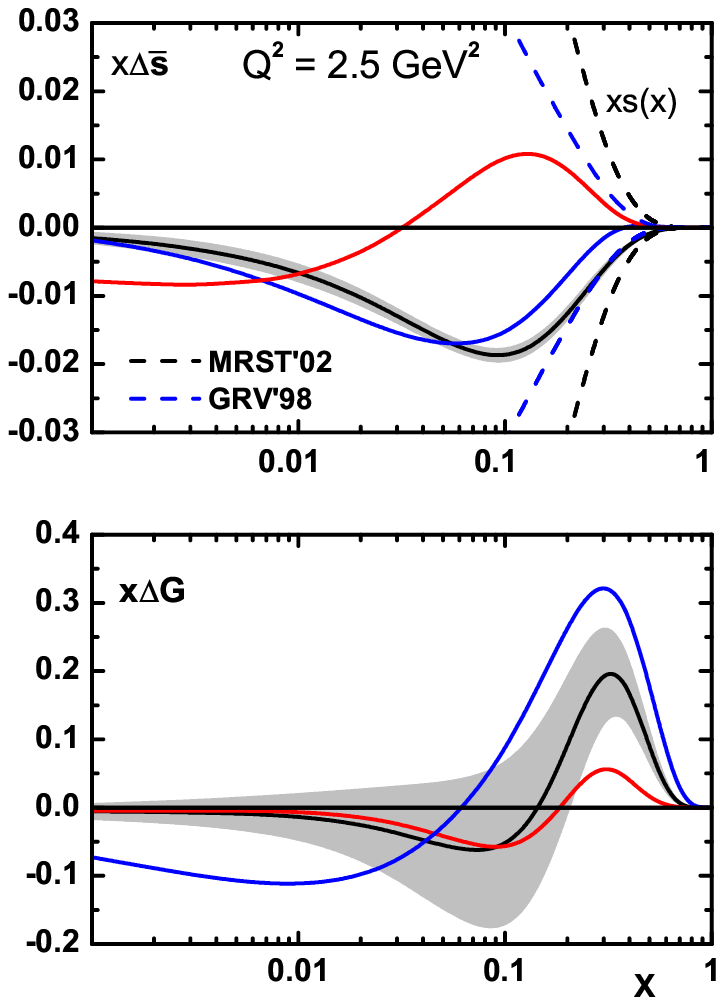}
  \caption{Comparison between LSS'06, AAC'08 and DSSV NLO PDFs
  in ($\rm \overline{MS}$) scheme.}
\end{figure}



\bibliographystyle{aipproc}   

\bibliography{sample}

\IfFileExists{\jobname.bbl}{}
 {\typeout{}
  \typeout{******************************************}
  \typeout{** Please run "bibtex \jobname" to optain}
  \typeout{** the bibliography and then re-run LaTeX}
  \typeout{** twice to fix the references!}
  \typeout{******************************************}
  \typeout{}
 }

\end{document}